\documentclass[twoside,12pt]{article}

\begin{document}

\thispagestyle{plain}
\setcounter{equation}{0}
\renewcommand{\thefootnote}{\fnsymbol{footnote}}

\begin{center}     \Large \bf
    Some effects of the quantum field theory   \\[2mm]
            in the early Universe
\end{center}

\vspace{1mm}
\begin{center}
{\bf A. A. Grib}{}\footnote{\small  E-mail: \  andrei\_grib@mail.ru }
\ and \
{\bf Yu. V. Pavlov}{}\footnote{\small E-mail: \ yuri.pavlov@mail.ru}
\end{center}
    \begin{center}           {\small  \it
A.\,Friedmann Laboratory for Theoretical Physics,   \\
30/32 Griboedov can, St.Petersburg, 191023, Russia   }
\end{center}

\vspace{1mm}
\begin{abstract}
    Effects of the vacuum polarization leading to change of the effective
gravitational constant and particle creation in the early Universe are
discussed.
Gauss-Bonnet type coupling to the curvature is considered.
Renormalization methods are generalized for such coupling
and the $N$-dimensional space-time.
Calculations of creation of entropy and visible matter of the Universe
by the gravity of dark matter are made.
\end{abstract}

\vspace{1mm}
\section{Introduction}
    Quantum field theory in curved space-time is the generalization of well
developed theory in Minkowski space-time, so it has some features of
the standard theory together with new ones due to the curvature.
  It was actively developed in the 70-ties of the last century
(see books~\cite{GMM1,GMM}) however some new results were obtained
just recently and in this paper we'll concentrate on these new results
as well as on some physical interpretation of the old ones.

   One of the main results of the theory for the early Friedmann Universe
was the calculation of finite stress-energy tensor for particle creation.
Let us begin from some remarks on what is meant by particle creation in
curved space-time?
    In spite of the absence of the standard definition of the particle
as the representation of the Poincar\'{e} group in curved space-time
where Poincar\'{e} group is not a group of motions, one still has
physically clear idea of the particle.
The particle is understood as the classical point like object moving along
the geodesic of the curved space-time.
This classical particle however is the quasiclassical approximation of some
quantum object, so the main mathematical problem is to find the answer
to the question: what is the Fock quantization of the field giving just
this answer for particles in this approximation?
The answer to this question was given by us in the 70-ties for conformal
massive scalar particles and spinor particles in Friedmann space-time
by use of the metrical Hamiltonian diagonalization method.
Creation of particles in the early Universe from vacuum by the gravitational
field means that for some small time close to singularity the stress-energy
tensor has the geometrical form, expressed through some combinations of
the Riemann tensor and its derivatives while for the time larger than the
Compton one it has the form of the dust of pointlike particles with mass
and spin defined by the corresponding Poincar\'{e} group representation.

    So our use of quantum field theory in curved space-time with its methods
of regularization (dimensional regularization and
Zeldovich-Starobinsky regularization) was totally justified by
the obtained results.
   However still unclear was the situation with
the minimally coupled scalar field where our method led to
infinite results for the density of created particles,
it is only recently that Yu.V.Pavlov~\cite{Pv} could find
the transformed Hamiltonian, diagonalization of which leads to finite
results for this case.

     The effect of particle creation in the early Universe can have
more general meaning concerning the origination of space-time
itself~\cite{Grib00}.
    It is reasonable to consider macroscopic Universe as classical
approximation of the quantum Universe.
    But what is quantum Universe?
    One can speculate that not only the metric but even the very existence
of space-time points described by some coordinates is due to existence
of operators of coordinates for some massive particles.
    In the ``empty'' quantum Universe there are no points!
    This is in some sense return to the Leibniz idea that space and time
are relations between particles and don't exist without them.
    It is important to note~\cite{Grib00} that space with space-like
intervals between particles  and time are necessary in order to
measure   or   have information for Boolean --  minded
classical observer about observables described by noncommuting operators.
    Noncommuting observables become commuting ones if one has some
``copying'' mechanism for the quantum system, when the same (identical)
system is observed in different points of space separated by the
space-like interval or at different moments of time.
    Time is needed for measuring noncommuting observables.
    This ``copying'' process makes possible contemplation of the quantum
Universe with its complementary characteristics by the classical observer.
    ``Copying'' is the same as particle creation.
    That is why particle creation can be considered as origination of
the classical Universe with its space and time from the quantum one.
    There is no necessity for quantization  of gravity from this point
of view, because space, time and metric are just some forms used by
the classical observer dealing with the quantum Universe.
    Entropy of the Universe arising due to such ``measurement'' process
is the other side of this process of particle creation and
origination of space-time~\cite{Grib00}.

     We use the system of units where $\hbar =c=1$.
     The signs of the curvature tensor and Ricci tensor are
chosen such that
$ R^{\, i}_{\ jkl} = \partial_l \, \Gamma^{\, i}_{\, jk} -
\partial_k \, \Gamma^{\, i}_{\, jl} +
\Gamma^{\, i}_{\, nl} \Gamma^{\, n}_{\, jk} -
\Gamma^{\, i}_{\, nk} \Gamma^{\, n}_{\, jl}  $,
$\ R_{ik} = R^{\, l}_{\ ilk}$,  $\ R=R_{\,l}^{\,l}$,
where $\Gamma^{\, i}_{\, jk}$ are Christoffel symbols.

\vspace{14pt}
\section{Some remarks on the physical interpretation of terms in vacuum
polarization dependent on mass}

    The calculations of the vacuum expectation value of the stress-energy
tensor of the quantized conformal massive scalar, spinor and vector
fields~\cite{GMM1,GMM} led to the expressions different for strong and weak
external gravitational field.
    By strong field one understands the field with the curvature much
larger than the one defined by the mass of the particle.
    For strong gravitational field it is basically polarization of vacuum
terms which become zero if the gravitation is zero, for weak gravitation
it is defined by particles created previously, so that if gravitation
becomes zero particles still exist.
    Vacuum polarization for strong gravitation consists of three terms:
the first is due to the conformal anomaly and it does not depend on mass
of the particle at all, the second is due to the Casimir effect and is
present for the closed Friedmann space, the third is vacuum polarization
dependent on mass of the particle which is for scalar conformal
particles~\cite{GMM1,GMM}
    \begin{equation}
\langle T_{ik} \rangle_m = \frac{m^2}{144 \pi^2}\,G_{ik} +
\frac{m^4}{64 \pi^2}\,g_{ik} \ln \biggl(\frac{\Re}{m^4}\biggr)\,.
\label{Tikm}
    \end{equation}
    Here $G_{ik}= R_{ik} - R g_{ik}/2 $ is the Einstein tensor,
$g_{ik}$ is the metrical tensor, $\Re$ is some geometrical term of
the dimension of the fourth degree of mass.
    It is interesting that the first term is the illustration of
the Sakharov's idea~\cite{Sakharov67D} of gravitation as the vacuum
polarization.
     If one takes the Planckean mass one has just
the standard expression for the term in the Einstein equation.
    One can also think that gravitation is the manifestation of
the vacuum polarization of all existing fields with different masses.
     However for weak gravitation for the time of evolution of the
Friedmann Universe larger than the Compton one defined by the mass there
is no Sakharov term, it is compensated.
    It is just manifestation of the fact, known also in quantum
electrodynamics that vacuum polarization is different in strong and weak
external fields being asymptotics of some general expression.
    So the physical meaning of this term in strong field is finite change
of the gravitational constant, so that the new effective gravitational
constant is different from that in the weak field, being
    \begin{equation}
\frac{1}{8 \pi G_{eff} } = \frac{1}{8 \pi G} +
\frac{m^2}{144 \pi^2} \,.
    \end{equation}
    So the larger is $m$, the smaller is $G_{eff}$.
    If $m$ is macroscopic~\cite{GribFrolov87}, then $ G_{eff} \to 0 $
which means some kind of ``asymptotic freedom'' for gravity if one
deals with macroscopic masses (like the mass of the Universe)
quantum mechanically.

    The second term on the right side~(\ref{Tikm}) describes what is now
called ``quint\-ess\-ence'' -- cosmological constant, dependent on time.
   This ``quintessence'' is different for different stages of
the evolution of the Friedmann Universe.
    We calculated it previously for the radiation
dominated Universe, so that together with the Sakharov term one has
    \begin{equation}
\langle T_{0}^{\,0} \rangle_m = -\frac{m^2}{48 \pi^2} \biggl( \frac{a'}{a^2}
\biggr)^2 \!- \frac{m^4}{16 \pi^2} \Biggl[ \ln(ma) + C + \frac{1}{4} +
\frac{1}{a^4} \int \limits_0^\eta \!d\eta_1 \frac{d a^2}{d\eta_1}
\int \limits_0^\eta \! d\eta_2 \frac{d a^2}{d\eta_2} \ln|\eta_1\!-\eta_2|
\Biggr].
    \end{equation}
    Here $a$ is the scale factor of the Friedmann space-time, $\eta $ is
the conformal time, $C=0.577 \ldots $ is the Euler constant.

\vspace{14pt}
\section{Creating of scalar particles in curved space-time}

   One of unsolved problems of quantum theory in curved space
is related to the definition of an elementary particle
and vacuum in a curved space-time.
   The reason is that in the case of a curved space  there is no group of
symmetries similar to the Poincar\'{e} group in the Minkowski space.
   If we believe that a particle is associated with a quantum of energy,
then according to quantum mechanics, the observation of particles at
an instant of time implies finding an eigenstate of the Hamiltonian.
    The Hamiltonian diagonalization method~\cite{GMM} takes this into
account automatically.

    The use of the Hamiltonian constructed from the metrical
energy-momen\-tum tensor, first proposed by A.A.Grib and
S.G.Mamayev~\cite{GribMamayev69} was successful in the homogeneous
isotropic space-time for the conformal scalar fields.
   But such Hamiltonian leads to the difficulties related to an
infinite density of created quasiparticles in the nonconformal
case~\cite{Fulling79}.
    The energy of such quasiparticles differs from the oscillator
frequency of the wave equation~\cite{CastF86,BMR98}.
    The nonconformal case is important for investigation by many reasons.
    Vector mesons and gravitons can be such particles.
    The additional nonconformal terms can be dominant in the vacuum
expectation values of the energy-momentum tensor~\cite{BLMPv}.
    In Ref.~\cite{Pv} the modified Hamiltonian was found so that the
density of the particles corresponding to its diagonal form and created
in the nonstationary homogeneous isotropic space-time is finite.
    In this section one gives the generalization of the method,
proposed in Ref.~\cite{Pv}, for the cases of conformally static
spaces and coupling with curvature of the general form
(see Ref.~\cite{PvIJA}).

    We consider a complex scalar field  $\varphi(x)$ of the mass $m$
with Lagrangian
    \begin{equation}
L(x)=\sqrt{|g|} \left[\, g^{ik}\partial_i\varphi^*\partial_k\varphi -
(m^2 + V_{\!g})\, \varphi^* \varphi \, \right],
\label{Lag}
\end{equation}
  and corresponding equation of motion
\begin{equation}
 (\nabla^i \nabla_{\! i} + V_{\!g} + m^2) \varphi(x)=0 \,,
\label{Eqm}
\end{equation}
   where ${\nabla}_{\! i}$ are the covariant derivatives in
$N$-dimensional space-time,
$ g\!=\!{\rm det}(g_{ik})$,  $\ V_{\!g}$ is the function of the
invariant combinations $g_{ik}$ and curvature tensor $R^{\,i}_{\ jkl}$.
    The equation~(\ref{Eqm}) is conformally invariant if $m=0$
and $V_{\!g}=\xi_c R $, where
$\xi_c = (N-2)/\,[\,4\,(N-1)] $ (conformal coupling).
   The case $V_{\!g}=0$ is the minimal coupling.
   In this section we investigate the case of conformally static metric
of the form
    \begin{equation}
ds^2 = dt^2 - a^2(t) \, \gamma_{\alpha \beta}({\bf x})\, dx^\alpha d x^\beta
= a^2(\eta) \left( d\eta^2 - \gamma_{\alpha \beta}({\bf x})\,
d x^\alpha d x^\beta \right),
\label{gikn}
\end{equation}
    where  $\alpha, \beta= 1, \ldots, N-1 $.
  It is realized, in particular, for the homogeneous isotropic space-time.
    The equation~(\ref{Eqm}) for the functions
$ \tilde{\varphi}=a^{(N-2)/2}\varphi $
in coordinates $(\eta, {\bf x})$ takes the form
   \begin{equation}
\tilde{\varphi}''- \Delta_{N-1} \tilde{\varphi} +
\biggl[ \left(m^2 + V_{\!g} \right) a^2 - \frac{N-2}{4} \left( 2c'+
(N-2) c^2 \right) \biggr] \, \tilde{\varphi}=0 \,,
\label{eft}
\end{equation}
    where the prime denotes the derivative with respect to
``conformal'' time $\eta$,
$\ c \equiv a'/a \ $,
$\ \Delta_{N-1} \equiv \gamma^{-1/2} \partial_\alpha (\sqrt{\gamma}\,
\gamma^{\alpha \beta} \partial_{\beta} ) $,
$\ \gamma = {\rm det}(\gamma_{\alpha \beta})$.
    Let us assume that Laplace-Beltrami operator $\Delta_{N-1}$
has the complete orthonormal set of the eigenfunctions
$\Phi_{\!J}({\bf x})$
    \begin{equation}
\Delta_{N-1} \Phi_{\!J}({\bf x}) = -\lambda^2(J)\, \Phi_{\!J}({\bf x}) \,,
\label{DelPhi}
\end{equation}
    where $J$ is set of $N-1$ indices (quantum numbers).
    The full set of the eq. (\ref{eft}) solutions can be found in the form
$ \tilde{\varphi}(x) = g_J(\eta) \Phi_{\!J} ({\bf x}) $\,,
  where
      \begin{equation}
g_J''(\eta) + \Omega^2(\eta)\,g_J(\eta)=0 \,,
\label{gdd}
\end{equation}
$\Omega$ is the oscillator frequency:
    \begin{equation}
\Omega^2(\eta)=\left( m^2 + V_{\!g} \right) a^2 -
\frac{N-2}{4} \left( 2c'+(N-2) c^2 \right) +\lambda^2(J) \,.
\label{Ome}
\end{equation}
    The expression
$ g_J(\eta)\,g_J^*{}'(\eta) - g_J^{\, \prime}(\eta)\,g_J^*(\eta)$
is the first integral of eq.~(\ref{gdd}).
    Further one normalizes the solutions of eq.~(\ref{gdd}) by condition
     \begin{equation}
g_J\,g_J^*{}' - g_J^{\, \prime}\,g_J^* = -2 i \,.
\label{norm}
\end{equation}
    For quantization, we expand the field $ \tilde{\varphi}(x) $
       \begin{equation}
\tilde{\varphi}(x)=\int \! d\mu(J)\,\biggl[ \tilde{\varphi}{}^{(+)}_J \,
a^{(+)}_J + \tilde{\varphi}{}^{(-)}_J \, a^{(-)}_J \,\biggr] \ ,
\label{fff}
\end{equation}
     where $d\mu(J)$ is the measure on the space of $\Delta_{N-1}$
eigenvalues,
    \begin{equation}
\tilde{\varphi}{}^{(+)}_J (x) =\frac{1}{\sqrt{2}}\,
g_J(\eta)\,\Phi^*_{\!J}({\bf x}) \ ,
\ \  \ \
\tilde{\varphi}{}^{(-)}_J (x)=\biggl(\tilde{\varphi}{}_J^{(+)}({x})
 \biggr)^* \,,
\label{fpm}
\end{equation}
     and impose the usual commutation relations on the operators
$ a_J^{(\pm)} , \ \stackrel{*}{a}\!{\!}_J^{(\pm)}$:
    \begin{equation}
\left[a_J^{(-)}, \ \stackrel{*}{a}\!{\!}_{J'}^{(+)}\right] =
\left[\stackrel{*}{a}\!{\!}_J^{(-)}, \ a_{J'}^{(+)}\right] =
\delta_{JJ'} \ , \ \ \
\left[a_J^{(\pm)}, \ a_{J'}^{(\pm)}\right] =
\left[\stackrel{*}{a}\!{\!}_J^{(\pm)}, \
\stackrel{*}{a}\!{\!}_{J'}^{(\pm)} \right]=0 \,.
\label{aar}
\end{equation}

    We construct Hamiltonian as canonical one for variables
$\tilde{\varphi}(x)$ and $\tilde{\varphi}^*(x)$.
  If we add $N$-divergence $(\partial J^i/ \partial x^i)$ to Lagrangian
density~(\ref{Lag}), where in the coordinate system $(\eta, {\bf x})$
the $ N$-vector
$\ (J^i)=(\sqrt{\gamma}\,c\,\tilde{\varphi}^*\, \tilde{\varphi}\,(N-2)/2,
\, 0, \, \ldots \,, 0) $, \
   the motion equation~(\ref{eft}) is invariant under this addition.
    By using the Lagrangian density
$ L^{\Delta}(x)=L(x)+({\partial J^i}/{\partial x^i})$, \
we obtain by integration on hypersurface $\Sigma :\ \eta=const $
of Hamiltonian density
$ h(x)= \tilde{\varphi}'\,(\partial L^{\Delta})/
(\partial \tilde{\varphi}') +
\tilde{\varphi}^{* \prime}\,(\partial L^{\Delta})/
(\partial \tilde{\varphi}^{* \prime})-L^{\Delta}(x) $,
the modified Hamiltonian
     \begin{eqnarray}
H(\eta) &=& \int_\Sigma h(x) \, d^{N-1}x = \int_\Sigma d^{N-1}x \,
  \sqrt{\gamma} \, \Biggl\{
\tilde{\varphi}^{* \prime} \tilde{\varphi}{}'
+\gamma^{\alpha \beta} \partial_\alpha\tilde{\varphi}^*
\partial_\beta \tilde{\varphi}+
\nonumber                \\
&+& \biggl[ \left(m^2+ V_{\!g} \right) a^2 - \frac{N-2}{4} \left(2c'+
(N-2)c^2\right)
\biggr]\,  \tilde{\varphi}^* \tilde{\varphi} \Biggr\} \,.
\label{Hp}
\end{eqnarray}
      Hamiltonian~(\ref{Hp}) is expressed in terms of the operators
$ a_J^{(\pm)} , \ \stackrel{*}{a}\!{\!}_J^{(\pm)}$ by
    \begin{equation}
H(\eta)= \! \int \! d\mu(J)  \biggl\{ \! E_J(\eta)\!
\left(\stackrel{*}{a}\!{\!}^{(+)}_J a^{(-)}_J +
\stackrel{*}{a}\!{\!}^{(-)}_{\bar{J}} a^{(+)}_{\bar{J}} \right) +
F_J(\eta)  \stackrel{*}{a}\!{\!}^{(+)}_J a^{(+)}_{\bar{J}} +
F^*_J(\eta) \stackrel{*}{a}\!{\!}^{(-)}_{\bar{J}} a^{(-)}_J \! \biggr\},
\label{H}
\end{equation}
    where
\begin{equation}
E_J(\eta)=\frac{1}{2} \biggl[\, |g_J'|^2+ \Omega^2 |g_J|^2 \,
\biggr]  \ ,
\ \ \ F_J(\eta)=\frac{\vartheta_{\!J}}{2} \biggl[\, g_J'{}^{\! 2} +
\Omega^2  g_J^{\, 2} \, \biggr]  \,,
\label{EJFJ}
\end{equation}
   here we used the normalization condition~(\ref{norm}) and the
special choice of the eigenfunctions so that for an arbitrary $J$ such
$\bar{J}$ exists, that $\Phi_{\!J}^*({\bf x}) = \vartheta_{\!J}
\Phi_{\!\bar{J}}({\bf x})$.
    In accordance with condition of the completeness such set can be
selected by the redefinition of eigenfunctions
$\Phi_{\!J}({\bf x})$.
    The Hamiltonian diagonalization for initial instant $\eta_0$
and normalization condition~(\ref{norm}) give
(under $\Omega^2(\eta_0)>0$):
    \begin{equation}
g_J^{\, \prime}(\eta_0)=i\, \Omega(\eta_0)\, g_J(\eta_0) \,, \ \
\ |g_J(\eta_0)|= \Omega^{-1/2}(\eta_0)\,.
\label{icg}
\end{equation}
    The Hamiltonian diagonalization for an arbitrary instant $\eta$
is realized in terms of the operators
$ \stackrel{*}{b}\!{\!}^{(\pm)}_J, \  b^{(\pm)}_J $
  related to the operators
$ \stackrel{*}{a}\!{\!}^{(\pm)}_J, \  a^{(\pm)}_J $
   via the time-dependent Bogoliubov transformations
\begin{equation}
a_J^{(-)}=\alpha^*_J(\eta) b^{(-)}_J(\eta) -
\beta_J(\eta) \vartheta_{\!J} b^{(+)}_{\bar{J}}(\eta), \ \ \
\stackrel{*}{a}\!{\!}_J^{(-)}=
\alpha^*_J(\eta) \! \stackrel{*}{b}\!{\!}^{(-)}_J\!(\eta) -
\beta_J(\eta) \vartheta_{\!J} \!
\stackrel{*}{b}\!{\!}^{(+)}_{\bar{J}}\!(\eta),
\label{db}
\end{equation}
     where the functions $\alpha_J, \ \beta_J $ satisfy
the initial conditions
$|\alpha_J(\eta_0)|=1, \ \beta_J(\eta_0)=0 $  and the identity
$ |\alpha_J(\eta)|^2-|\beta_J(\eta)|^2=1 $.
   Substituting expansion~(\ref{db}) in  (\ref{H})
and requiring that the coefficients of the nondiagonal terms
$ \stackrel{*}{b}\!{\!}^{(\pm)}_J  b^{(\pm)}_J $
vanish, we obtain
    \begin{equation}
-2\alpha_J\beta_J \vartheta_{\!J}E_J + \alpha_J^2 F_J
+\beta_J^2 \vartheta_{\!J}^2 F_J^* = 0   \,, \ \ \
|\beta_J|^2=
\frac{1}{4\Omega}\biggl( |g_J'|^2+\Omega^2 |g_J|^2 \biggr) -\frac{1}{2} \,,
\label{bFJ}
\end{equation}
     \begin{equation}
H(\eta) =\int \! d\mu(J) \,\Omega(\eta)
\left(\,\stackrel{*}{b}\!{\!}^{(+)}_J b^{(-)}_J +
\stackrel{*}{b}\!{\!}^{(-)}_{\bar{J}} b^{(+)}_{\bar{J}} \right).
\label{Hbb}
\end{equation}
   Therefore, the energies of quasiparticles corresponding to the diagonal
form of the Hamiltonian~(\ref{H}) are equal to the oscillator frequency
$\Omega(\eta)$.

   The vacuum state for the instant $\eta$ is defined by
$ b^{(-)}_J(\eta)| 0_\eta \rangle \!=
\stackrel{*}{b}\!{\!}^{(-)}_J(\eta)| 0_\eta \rangle{=0}.$
    The state $| 0 \rangle = |0_{\eta_0} \rangle $ contains
$|\beta_J(\eta)|^2$ quasiparticle pairs corresponding to the operators
$b^{(\pm)}_J(\eta)$ in every mode~\cite{GMM,BD}.
    The density of created particles is proportional to
$\int\! d\mu(J)\, |\beta_J|^2 $.
     The function $ S(\eta)=|\beta_J(\eta)|^2 $ obeys the integral
equation
    \begin{equation}
S(\eta)=\frac{1}{2}\,\int_{\eta_0}^\eta d\eta_1 \,
w(\eta_1)\, \int_{\eta_0}^{\eta_1} \! d\eta_2 \,w(\eta_2)\,
(1+2 S(\eta_2)) \cos[2\,\Theta(\eta_2,\eta_1)]  \,,
\label{iuS}
\end{equation}
    where
$$
w(\eta)=\Omega'(\eta)/\,\Omega(\eta) \ , \ \ \ \
\Theta(\eta_2, \eta_1)=\int_{\eta_2}^{\eta_1} \! \Omega(\eta)\,d \eta \,.
$$
    Using the first iteration and taking into account that
$ \Theta(\eta_2, \eta_1) \to \lambda (\eta_1 - \eta_2) $ in
$\lambda \to \infty $
    one can represent~(\ref{iuS}) as
\begin{equation}
S(\eta) \approx \frac{1}{4} \left| \int_{\eta_0}^\eta
w(\eta_1) \exp( 2 i \lambda \eta_1 )\, d\eta_1 \right|^{\,2}.
\label{1it}
\end{equation}
     Consequently, one can see, that
$S \sim \lambda^{-6} $ in  $\lambda(J) \to \infty $.

    Therefore, in 4-dimensional space-time with metric~(\ref{gikn})
the density of created particles is finite.
    So it is shown differently from~\cite{Fulling79,BD}, that even for
minimal coupling the result is finite!
    From eq.~(\ref{iuS}) for $S(\eta)$ one can see
that under $ V_{\!g}=\xi R $ the densities of scalar particles
created in Friedmann radiative-dominant Universe don't depend on $\xi$.
    So the results will be for  $t \gg m^{-1}$  the same as for
conformal coupled particles obtained earlier~\cite{GMM1,GMM}
(see eq.~(\ref{NbM}) in this paper).

\vspace{14pt}
\section{Scalar field with Gauss-Bonnet type coupling to the curvature}

    Usually for scalar field in curved space-time one writes
the eq.~(\ref{Eqm}) with $V_{\!g}=\xi R$.
    The condition $\xi=\xi_c$ consistent with conformal invariance
of the equations for massless field is considered sometimes as preferable
(see for example~\cite{SonegoFaraoni93,GribP}).
    For scalar fields in inflation models~\cite{Linde} minimal
coupling $\xi=0$ usually is used.
    In general case one can take arbitrary values of $\xi$
\cite{BLMPv,HMPM00}.
    Renormalization of interacting fields in curved space-time is
inconsistent with conformal invariance not only of the effective action
but also of usual action~\cite{BD}.
    Models with arbitrary $\xi$ without conformal invariance on
the classical level can be generalized by adding of the quadratic in
curvature terms
    \begin{equation}
V_{\!g} = \xi R + \zeta R^2 + \kappa R_{ij}R^{ij} +
\chi R_{ijkl}R^{ijkl} + \ldots
\label{VRR}
\end{equation}
    Nonzero coupling constants in these terms in (\ref{VRR})
having the dimension (mass)${}^{-2}$ lead usually to derivatives of
the third and fourth order in the metrical energy-momentum tensor (EMT)
and in Einstein equations.
    As it is known~\cite{Wald77} such terms with higher derivatives
even for small coefficients lead to the radical change of the theory.
    If one has the condition that metrical EMT of the scalar field
does not contain derivatives of metric higher than the second order,
one can take for $V_{\!g}$
    \begin{equation}
V_{\!g} = \xi R + \zeta R_{GB}^{\,2} \,,  \ \ \  \mbox{where} \ \ \ \ \
R_{GB}^{\,2} \stackrel{\rm def}{=}
R_{lmpq} R^{\,lmpq} - 4 R_{lm} R^{\,lm} + R^2\,.
\label{V}
\end{equation}
    On even-dimensional spaces due to the Gauss-Bonnet
theorem \cite{TsFKS} Euler characteristic of compact orientable
manifold $M$ with Riemann metric is equal to
       \begin{equation}
\chi(M) = \int_M E(x) \sqrt{ g } \, d^N x  \,,
\label{chi}
\end{equation}
    where $E(x) = - (4 \pi)^{-1} R(x) $ for $N=2$\
and $E(x) = R_{GB}^{\,2}/(128 \pi^2) $  for $N=4$.
    So the case with $V_{\!g}$ defined by~(\ref{V}) can be called
Gauss-Bonnet type coupling.

    The condition of absence of higher derivatives of metric was
considered earlier~\cite{Lovelock71} as basic for multidimensional
generalizations of the gravity theory.
    It was shown that the demands of symmetry in indices,
covariant conservation and
the condition that the tensor generalizing Einstein tensor $G_{ik}$
is constructed from $g_{ik}$ and only its first and second
derivatives give the Einstein equations with cosmological
constant in dimension four.
    But in dimension larger than four the corresponding equations
can be obtained from the action which is a sum of contributions
associated with continuation to this dimension of Euler characteristics
of all lower even dimensions~\cite{Lovelock71}.
    The demand of absence of ghosts in low energy approximations
of string theories gives the Einstein-Gauss-Bonnet gravity with
$R_{GB}^{\,2}$ density in action for higher dimensions~\cite{Zwiebach85}.
    The Gauss-Bonnet coupling of scalar field with gravity
was also considered in dilaton theories
(see for example~\cite{MignemiStewart93}).

    Taking variational derivatives of the action for scalar field with
coupling~(\ref{V}) one obtains the following expression for metrical EMT:
     \begin{eqnarray}
T_{ik} \! &\!=\!& \! \partial_i\varphi^* \partial_k\varphi+
\partial_k\varphi^*\partial_i\varphi - g_{ik} \partial^{\,l}\varphi^*
\partial_l\varphi + g_{ik} m^2 \varphi^* \varphi -
     \nonumber    \\
\! &\!-\!& \! 2 \xi \left( G_{ik} + \nabla_{\! i} \nabla_{\! k} -
g_{ik} \nabla^l \nabla_{\! l} \right) (\varphi^* \varphi)  -
2 \zeta \left( E_{ik} + P_{ik} \right) (\varphi^* \varphi) \,,
\label{TGB}
\end{eqnarray}
    where
\begin{eqnarray}
E_{ik} &=& \frac{\delta  {\displaystyle \int \!  R_{GB}^{\,2} \,
\sqrt{|g|} \, d^N x}}{\sqrt{|g|}  \, \delta g^{ik}} =
2 R_{ilmp} R_k^{\ lmp} -\frac{g_{ik}}{2} R_{lmpq} R^{\,lmpq} -
\nonumber        \\
&-& 4 R^{\, lm} R_{limk} -4 R_{il} R_k^{\,l} + 2 g_{ik} R_{lm} R^{\,lm}
+ 2 R R_{ik} - \frac{g_{ik}}{2} R^2 =
\label{Eik}     \\
&=& 2 C_{ilmp} C_k^{\ lmp} - \frac{g_{ik}}{2} C_{lmpq} C^{lmpq}
- (N - 4) {}^{(3)}\!H_{ik} \,,    \nonumber
\end{eqnarray}
      \begin{eqnarray}
P_{ik} &=& 2 \, \biggl[ \, R \nabla_{\! i} \nabla_{\! k} +
2 R_{ik} \nabla_{\! l} \nabla^l + 2 g_{ik} R_{lm} \nabla^l \nabla^m -
g_{ik} R \nabla_{\! l} \nabla^l -
\phantom{xx}      \nonumber     \\
&-& 4 R_{l(i} \nabla_{\! k)} \nabla^l -
2 R_{ilkm} \nabla^l \nabla^m \biggr] =
- 4 C_{ilkm} {\nabla}^l {\nabla}^m +   \phantom{xxxx}
\label{Pik}   \\
&+& 4 \, \frac{N\!-\!3}{N\!-\!2} \Biggl( R_{ik}
{\nabla}_{\! l} \nabla^l + g_{ik} R_{lm} {\nabla}^l \nabla^m -
\nonumber  \\
&-& 2 R_{l(i} {\nabla}_{\! k)} \nabla^l +
\frac{ N R }{2(N\!-\!1)} \left( {\nabla}_{\! i} \nabla_{\! k}
- g_{ik} {\nabla}_{\! l} {\nabla}^l \right) \Biggr)\,,  \nonumber
\end{eqnarray}
      \begin{equation}
{}^{(3)}\!H_{ik} = \frac{4 C_{ilkm} R^{\,lm}}{N-2} +
\frac{2(N\!-3)}{(N\!-2)^2} \left[ 2 R_{il} R_k^{\,l} -
\frac{N R R_{ik}}{N-1} - g_{ik} \Biggl( R_{lm} R^{\,lm} -
\frac{(N\!+\!2) R^2}{4(N\!-1)} \Biggr) \right],
\label{3Hik}
\end{equation}
      \begin{equation}
C_{iklm} = R_{iklm} + \frac{2}{N\!-\!2} \biggl( R_{m\,[\,i}\,
g_{k\,]\,l} - R_{l\,[\,i}\, g_{k\,]\,m} \biggr) + \frac{2 \, R
\,g_{l\,[\,i}\,g_{k\,]\,m} }{(N\!-\!1)(N\!-\!2)} \,,
\label{Ciklm}
\end{equation}
    parentheses (square brackets) in indices mean
symmetrization (antisymmetrization).
    Expressions for EMT using conformal Weyl tensor
$C_{iklm}$ are convenient for calculation as in conformally
flat ($C_{iklm}=0$), for example homogeneous isotropic spaces,
as in Ricci flat ($R_{ik}=0$) spaces.
    For $ N=2,3 $ one has $R_{GB}^{\, 2} \equiv 0$ and there are no new
effects due to $\zeta \ne 0$.
    In 4-dimensional space-time $E_{ik}=0$ (see Ref.~\cite{Lanczos38}),
but $P_{ik}(\varphi^*\varphi) \ne 0$ for general metric and
$\varphi(x) \ne const$.

    Vacuum averages of EMT for quantized fields are divergent.
    For the analysis of the geometrical structure of divergences of
vacuum averages of EMT one can use dimensionally regularized
effective action.
    For complex scalar field with eq.~(\ref{Eqm}) in loop approximation
the effective action
can be written~\cite{Bunch79,Fulling} as
       \begin{equation}
S_{eff} = \int L_{eff}(x) \sqrt{|g|}\, d^N x \,,
\label{Seff}
\end{equation}
   where
       \begin{equation}
L_{eff}(x) = (4 \pi)^{-N/2} \left( \frac{M}{m}
\right)^{2\varepsilon} \sum_{j=0}^\infty a_j(x) \, m^{N_0-2j}\,
\Gamma\biggl(j-\frac{N}{2}\biggr)  \,,
\label{Leff}
\end{equation}
       \begin{equation}
a_0(x)=1 \,, \ \ \ \ \
a_1(x) = \frac{1}{6}\,R - V_{\!g} = \left( \frac{4-N}{12 (N \!-\!1)} +
\Delta \xi \right) R - \zeta R_{GB}^{\,2} \,,
\label{a0a1}
\end{equation}
       \begin{equation}
a_2(x) = \frac{R_{lmpq} R^{\,lmpq}}{180}  -
\frac{R_{lm} R^{\,lm}}{180}  + \frac{R^2}{72}  -
\frac{\nabla^l \nabla_{\!l} R}{30}  -
\frac{R V_{\!g}}{6}  + \frac{V_{\!g}^2}{2}  +
\frac{\nabla^l \nabla_{\!l} V_{\!g}}{6}  \,,
\label{a2}
\end{equation}
    $N$ is space-time dimension which is taken as some variable,
analytically continuated in complex plane,
$\varepsilon$ is some complex parameter, $M$ is some constant of the
dimension of mass~\cite{tHooft}, necessary for correct dimension of
$L_{eff}$, (length)${}^{-N_0}$ for $ N = N_0-2\varepsilon $,
$ \ \Gamma(z)$ is gamma-function, \ $  \Delta \xi \equiv \xi_c - \xi $.
     Using~(\ref{Ciklm}) for Gauss-Bonnet coupling~(\ref{V})
one can write $a_2$ as
    \begin{eqnarray}
a_2(x) =
\left( \frac{ (N\!-\!4) (N\!-\!6)}{480\,(N\!-\!1)^2} - \frac{\Delta \xi
(N\!-\!4)}{12\,(N\!-\!1)} +\frac{(\Delta\xi)^2}{2} \right)\! R^2
+ \frac{(N\!-\!2) C_{lmpq} C^{\,lmpq}}{240\, (N - 3)} +
\nonumber        \\
{}+ \frac{(N\!-\!6)\, R_{GB}^{\,2}}{720\, (N\!-\!3)} +
\left( \frac{(N\!-\!4)}{12(N\!-\!1)} - \Delta \xi  \right)
\zeta R R_{GB}^{\,2}  + \frac{\zeta^2 R_{GB}^{\,4}}{2}
- \frac{1}{6} \nabla^l \nabla_{\!l} \biggl( \frac{1}{5} R - V_{\!g} \biggr)
\,.    \label{a2m}
\end{eqnarray}

    First $[N_0/2]+1$ terms in~(\ref{Leff}) are excluded to get
the renormalized $L_{eff}$
($[b]$ denotes the integer part of the number $b$).
    By variation in $g_{ik}$ of terms $j=0,1,2$ in effective action
one obtains terms subtracted from vacuum EMT
     \begin{equation}
T_{ik,\varepsilon}[0]=- \frac{m^{N_0}}{2^{N_0} \pi^{N_0/2}} \left(
\frac{4 \pi M^2}{m^2} \right)^{\! \displaystyle \varepsilon }
\Gamma  \biggl( \varepsilon - \frac{N_0}{2} \biggr) \, g_{ik} \,,
\label{TE0}
\end{equation}

\newpage
    \begin{eqnarray}
T_{ik,\varepsilon}[1] &=& \frac{m^{N_0-2}}{2^{N_0-1} \pi^{N_0/2}}
\left( \frac{4 \pi M^2}{m^2} \right)^{\! \displaystyle \varepsilon }
\Gamma \biggl( 1 -  \frac{N}{2} \biggr) \left[
\left( \frac{1}{6} - \xi \right) G_{ik} -\zeta E_{ik} \right] =
\nonumber    \\
&=& \frac{m^{N_0-2}}{2^{N_0-1} \pi^{N_0/2}} \!
\left(\frac{4 \pi M^2}{m^2} \! \right)^{\! \displaystyle \varepsilon }
\times   \label{TE1} \\
&\times& \!
\Biggl[ \Delta \xi\, \Gamma \biggl(1\! - \frac{N}{2} \biggr) G_{ik} -
\frac{\Gamma\! \left(3 - \frac{N}{2} \right)}{(N-2)}
\Biggl(  \frac{G_{ik}}{3 (N\!-\!1)} + \frac{ 4 \zeta E_{ik}}{(N\!-4)}
\Biggr) \Biggr],   \nonumber
\end{eqnarray}
      \begin{eqnarray}
T_{ik,\varepsilon}[2] \!=\!
\frac{m^{N_0-4}}{(4\pi)^{\frac{N_0}{2}}}
\!\left( \!\frac{4 \pi M^2}{m^2}\!\right)^{\! \displaystyle \varepsilon }
\!\left\{ \frac{\Gamma \left( 2 - \frac{N}{2}\right)}{360 (N \!-\! 3) }
\biggl[ (N\!-\!6) E_{ik} + 3 (N\!-\!2) W_{ik} \biggr] \right. +
 \nonumber                \\
{} + \Biggl[ \, \frac{\Gamma \left( 4 - \frac{N}{2}\right)}{60
\, (N - 1)^2 } + \Delta \xi\, \frac{\Gamma \left(3 - \frac{N}{2}
\right)}{3\, (N - 1) } + (\Delta \xi)^2  \, \Gamma \biggl( 2 -
\frac{N}{2} \biggr) \,\Biggr] {}^{(1)}\! H_{ik} -
\nonumber           \\
{} - \zeta \, \frac{\Gamma \left(3\!-\!\frac{N}{2}\right)}{3(N\!-\!1)}
\Biggl[ \biggl( R_{ik} + {\nabla}_{\! i} {\nabla}_{\! k} -
g_{ik} {\nabla}_{\! l}{\nabla}^l \biggr) R_{GB}^{\, 2}  +
\biggl( E_{ik} + P_{ik} \biggr) R
\Biggr] +
      \label{TE2ik}     \\
{}+ \zeta \, 2 \, \Gamma \left( 2 - \frac{N}{2}\right)
\Biggl[ \, \zeta\, \biggl( \frac{g_{ik}}{4} R_{GB}^{\, 2} + E_{ik} +
P_{ik} \biggr) \, R_{GB}^{\, 2} -
   \nonumber           \\
{} - \Delta \xi \, \Biggl( \biggl(R_{ik} + {\nabla}_{\! i} {\nabla}_{\! k}
- g_{ik} {\nabla}_{\! l}{\nabla}^l \biggr) R_{GB}^{\, 2}  +
\biggl( E_{ik} + P_{ik} \biggr) R \Biggr) \Biggr] \Biggr\} \,,
\nonumber
\end{eqnarray}
    where
    \begin{equation}
{}^{(1)}\! H_{ik}= \frac{\delta  {\displaystyle \int \!  R^2  \,
\sqrt{|g|} \, d^N x}} {\sqrt{|g|} \,  \delta g^{ik}} = 2 \biggl(
\nabla_{\! i} \nabla_{\! k} R - g_{ik} \nabla^l \nabla_{\! l} R
\biggr) + 2 R \biggl( R_{ik} -\frac{1}{4} R \, g_{ik} \biggr),
\label{1Hik}
\end{equation}
    \begin{eqnarray}
W_{ik} = \frac{\delta
{\displaystyle \int \!  C_{lmpq} C^{\,lmpq} \sqrt{|g|} \, d^N x}}
{\sqrt{|g|}  \, \delta g^{ik}} = \!
E_{ik} \!+\! \frac{4(N\!\!-\!3)}{N-2}\! \Biggl( 2 R^{ lm} R_{limk} \!-\!
\frac{ g_{ik} }{ 2 } R_{lm} R^{ lm}\! -
\nonumber        \\
- \frac{ N R R_{ik} }{ 2(N\!-\!1) } +
\frac{ N R^2 g_{ik} }{ 8(N\!-\!1) } +
\frac{N-2}{2(N\!\!-\!1)} \nabla_{\! i} \nabla_{\! k} R +
\frac{ g_{ik} }{ 2(N\!\!-\!1) } \nabla^l \nabla_{\! l} R - \!
\nabla^l \nabla_{\! l} R_{ik} \Biggr).
\label{Wik}
\end{eqnarray}
     In conformally flat case
$ C_{iklm}\!=0 $,\ \ $E_{ik}/(4\!-\!N)\!={}^{(3)}\!H_{ik} $,\ \
$W_{ik}\!=0$  and so
    \begin{equation}
T_{ik,\varepsilon}[1] \!=\!
\frac{m^{N_0-2}}{2^{N_0-1} \pi^{N_0/2}}\!
\left(\! \frac{4 \pi M^2}{m^2} \! \right)^{\!\! \displaystyle \varepsilon }
\! \Biggl[ \Delta \xi\, \Gamma \biggl(1 \!- \frac{N}{2} \biggr) G_{ik}
+ \frac{\Gamma\! \left(\!3 - \frac{N}{2} \right)}{(N\!-2)}
\Biggl(\! \zeta 4 {}^{(3)}\!H_{ik} - \frac{G_{ik}}{3 (N\!-\!1)}
\Biggr) \Biggr],
\label{TE1kp}
\end{equation}

\newpage
      \begin{eqnarray}
T_{ik,\varepsilon}[2] = \frac{m^{N_0-4}}{(4\pi)^{N_0/2}}
\left( \frac{4 \pi M^2}{m^2}\right)^{\! \displaystyle \varepsilon }
\left\{ {}^{(3)}\! H_{ik} \left[
\frac{ - \Gamma \left(4 -\frac{N}{2}\right) }{ 90\, (N-3) } \right. +
\right.
\phantom{xxxx}   \nonumber      \\
{} + \left. \zeta 4 \Gamma \left(3 -\frac{N}{2}\right)\!
\left( \Biggl( \frac{N-4}{12(N\!-\!1)} - \Delta \xi \Biggr) R +
\zeta R_{GB}^{\,2} \right) \right] +
\phantom{xxxx}    \nonumber           \\
{} +  {}^{(1)}\! H_{ik} \Biggl[ \,
\frac{\Gamma \left( 4 - \frac{N}{2}\right)}{60\, (N - 1)^2 } +
\Delta \xi\, \frac{\Gamma \left(3 - \frac{N}{2} \right)}{3\, (N - 1) }
+ (\Delta \xi)^2  \, \Gamma \biggl( 2 - \frac{N}{2} \biggr) \,\Biggr] -
\phantom{x}    \label{TE2kp}      \\
{} - \zeta \, \frac{\Gamma \left(3\!-\!\frac{N}{2}\right)}{3(N\!-\!1)}
\Biggl[ \biggl( R_{ik} + {\nabla}_{\! i} {\nabla}_{\! k} -
g_{ik} {\nabla}_{\! l}{\nabla}^l \biggr) R_{GB}^{\, 2}  + P_{ik} R
\Biggr] + \zeta \, 2 \, \Gamma \left( 2 \!-\! \frac{N}{2}\right)
\times     \nonumber    \\
{} \times \Biggl[ \zeta \biggl( \frac{g_{ik}}{4} R_{GB}^{\, 2} +
P_{ik} \biggr) R_{GB}^{\, 2} -
\Delta \xi \Biggl( \biggl(R_{ik} + {\nabla}_{\! i} {\nabla}_{\! k}
- g_{ik} {\nabla}_{\! l}{\nabla}^l \biggr) R_{GB}^{\, 2} +
P_{ik} R \Biggr) \Biggr] \Biggr\}.
\nonumber
\end{eqnarray}
     Supposing that vacuum averages of EMT, i.e.
$\langle \, T_{ik}\, \rangle$, are sources of the gravity
field~\cite{GMM,BD}
     \begin{equation}
G_{ik} + \Lambda g_{ik} = - 8 \pi G
\biggl( T_{ik}^{\,b} + \langle \, T_{ik}\, \rangle \biggr)\,,
\label{Eeq}
\end{equation}
   where $\Lambda$, $G$ are cosmological and gravitational constants,
$ T_{ik}^{\,b} $ is EMT of the background,
and taking into account eqs. (\ref{TE0})--(\ref{TE2ik})
one comes to the following conclusion.
    First three subtractions from the vacuum EMT in $N$-dimensional
space-time correspond to renormalization of the cosmological and
gravitational constants and parameters in quadratic, cubic and
fourth degrees in curvature terms in the bare gravitational
Lagrangian
     \begin{equation}
L_{gr,\,\varepsilon}=\sqrt{|g|} \left[
\frac{ R \!-\! 2 \Lambda_\varepsilon }{16 \pi G_\varepsilon} +
\alpha_\varepsilon  R_{GB}^{\,2} +
\beta_\varepsilon   R^2 +
\gamma_\varepsilon  C_{lmpq} C^{lmpq} +
\delta_\varepsilon  R R_{GB}^{\,2} +
\theta_\varepsilon  R_{GB}^{\,4}     \right].
\label{Lgr0}
\end{equation}
     Subtraction $ T_{ik,\varepsilon}[0] $ due to (\ref{TE0}),
leads to infinite renormalization of the cosmological
constant $\Lambda_\varepsilon$.
    From~(\ref{TE1}) it follows that subtraction $ T_{ik,\varepsilon}[1] $
correspond to renormalization of the gravitational constant
$G_\varepsilon$ (finite for $N \to 4$, $\xi=\xi_c$)
and infinite renormalization of $\alpha_\varepsilon$ (for $\zeta \ne 0$).
    Subtraction $ T_{ik,\varepsilon}[2] $ due to
(\ref{a2m}) and (\ref{TE2ik}) corresponds to renormalization
of the parameters
$\alpha_\varepsilon, \beta_\varepsilon, \gamma_\varepsilon,
\delta_\varepsilon, \theta_\varepsilon$.
    For $\xi=\xi_c$ and $N \to 4$, the parameters
$\beta_\varepsilon$, $\delta_\varepsilon$ have finite renormalization.

     Note that introduction of the term $\alpha_\varepsilon  R_{GB}^{\,2}$
in ~(\ref{Lgr0}) is consistent with dimensional regularization
($E_{ik} \equiv 0$ only for integer $N=2,3,4$).

     For $N\to 4$ the products $E_{ik}\Gamma (1-(N/2))$ and
$E_{ik}\Gamma (2-(N/2))$ have finite limits for arbitrary metric
of space-time because the dependence of $E_{ik}$ on $N$
is fraction-rational in analytic continuation on dimension,
$E_{ik}=0$ for $N=4$ and gamma-function has pole of the first order in
points $0$, $-1$.
    So the corresponding terms in~(\ref{TE1}), (\ref{TE2ik})
are finite and there is no necessity for subtracting them to get
finite terms in~(\ref{Eeq}) leading to back reaction of
the quantized field in metric.
    However the effective action is divergent without such
subtractions and the anomalous trace of vacuum EMT is different from
the standard even for $\zeta=0$.
    Due to the appearance of such terms for different regularization
procedures~\cite{GMM,BD} it is convenient to have them in
counterterms in vacuum EMT.

    Values of renormalized parameters are fixed by experiment
and it can be~\cite{GMM1,GMM} that renormalized parameters for
non-Einstein terms in the gravity Lagrangian are zero.

    Now consider calculating of the renormalized vacuum EMT for
$N$-dimensional quasi-Euclidean space-time with metric
    \begin{equation}
ds^2 = d t^2 - a^2(t)\, d {\bf x}^2 =
a^2(\eta) \left( d{\eta}^2 - d {\bf x}^2 \right).
\label{gik}
\end{equation}
    In this case one can choose
\begin{equation}
\Phi_J({\bf x}) = (2 \pi)^{-(N-1)/2}\, e^{-i \lambda_\alpha x^\alpha} \,,
\label{Phi}
\end{equation}
$$
J = \{ \lambda_1, \ldots , \lambda_{N-1} \} \,, \ \
- \infty < \lambda_\alpha < +\infty \,, \ \ \
\lambda = \sqrt{\sum \limits_{\alpha=1}^{N-1} \lambda_\alpha^2} \ ,
\ \ \ g_J(\eta) = g_\lambda(\eta) \,.
$$
    It is convenient to express vacuum averages of the EMT operator
for Fock vacuum $| 0\rangle $ annihilated by operator
$a_J^{(-)}\!,\ \stackrel{*}{a}\!{\!}_{J}^{(-)}$~(\ref{aar})
through linear combinations of functions~$g_\lambda $, $g_\lambda^*$:
    \begin{equation}
S=\frac{|g_\lambda'|^2 + \Omega^2\,|g_\lambda|^2}{4 \, \Omega}
-\frac{1}{2} \,, \ \ U=\frac{\Omega^2 \, |g_\lambda|^2-
|g_\lambda'|^2}{2\, \Omega} \,  \,, \ \ V= - \frac{d (g_\lambda^*
g_\lambda)}{2\, d \eta} \,,
\label{SUV}
\end{equation}
    which due to (\ref{gdd}) satisfy differential equations
    \begin{equation}
S'= \frac{\Omega'}{2\, \Omega} \, U \ , \ \ \ U'=
\frac{\Omega'}{\Omega} \, (1+ 2 S) - 2 \, \Omega V \ , \ \ \ V'= 2
\,\Omega \, U \,.
\label{sdu}
\end{equation}
    Taking initial conditions
$ S(\eta_0)=U(\eta_0)=V(\eta_0)=0$ following from~(\ref{icg})
the eqs.~(\ref{sdu}) can be written as integral equations of the Volterra
type~(\ref{iuS}) and
     \begin{equation}
U(\eta) + i V(\eta) = \int_{\eta_0}^{\, \eta} \! w(\eta_1)\, (1+2
S(\eta_1)) \exp[ 2\,i\,\Theta(\eta_1,\eta)]\,d\eta_1  \,.
\label{ie1}
\end{equation}
    Substituting expansion (\ref{fff}) in eq. (\ref{TGB}) and using
(\ref{aar}), (\ref{Eik})--(\ref{3Hik}),  (\ref{SUV}),
one obtains the following (diverging) expression for vacuum EMT averages
    \begin{equation}
\langle 0 |\,T_{ik}| 0\rangle =\frac{B_N}{a^{N-2}} \int_0^\infty \!
 \tau_{ik} \lambda^{N-2} d \lambda \,,
\label{Ttik}
\end{equation}
    where
$ B_N\!=\!\left[ 2^{N-3}\pi^{(N-1)/\,2} \, \Gamma((N\!-\!1)/2) \right]^{-1}$,
           \begin{equation}
\tau_{00}=\Omega \biggl( S+\frac{1}{2}\biggr) + (N\!-\!1)
\biggl( \Delta \xi - \tilde{\zeta} c^2  \biggr)
\left[ c V + \biggl( c'+(N \!-\! 2) c^2 \biggr)
\frac{1}{\Omega} \left(\!S+\frac{1}{2} U + \frac{1}{2} \right) \right],
\label{ts00}
\end{equation}
    \begin{eqnarray}
\tau_{\alpha \beta} &=& \delta_{\alpha \beta} \, \Biggl\{\,
\frac{1}{(N \!-\! 1) \Omega}  \left[ \lambda^2 \left( S +
\frac{1}{2} \right) - \left( \Omega^2 - \lambda^2 \right)
\frac{U}{2} \, \right] +
       \nonumber   \\
&+& \biggl[ \Delta \xi (N\!-\!1) - \tilde{\zeta} \biggl( (N\!+\!1)
c^2 - 2 c'\biggr) \biggr] c V -
2 \biggl( \Delta \xi \!- \tilde{\zeta} c^2 \biggr) \Omega U -
     \label{tsab}   \\
&-& \frac{1}{\Omega} \left( S+\frac{1}{2} U + \frac{1}{2} \right)
\Biggl[ \Delta \xi (N\!-\!1) c'- \tilde{\zeta} (N\!-\!1) c^2
\! \left( 3c'\!-\! 2 c^2 \right) \Biggr] \Biggr\},
\nonumber
\end{eqnarray}
  $ \tilde{\zeta} \equiv \zeta a^{-2}\, 2 (N\!-\!2)(N\!-\!3)$.

   The vacuum expectation~(\ref{Ttik}) has $[N/2]+1 $ different types
of divergences:
$ \sim \lambda^N, \ \lambda^{N-2}, \ldots , \ln \lambda $ \ \
if $N$ is even, and
$\ \sim \lambda^N, \ \lambda^{N-2}, \ldots , \lambda $ \ \
if $N$ is odd.
   For renormalization we are using $n$-wave procedure
proposed by Zeldovich and Starobinsky~\cite{ZlSt}
    \begin{equation}
\langle 0|\,T_{ik}| 0\rangle_{ren} = \frac{B_N}{a^{N-2}}
\int \limits_{0}^{\infty} \! \lambda^{N-2} \, \biggl[ \tau_{ik} -
\sum \limits_{l=0}^{[N/2]} \tau_{ik}[l]\,\biggr]\,d\lambda \,,
\label{Trik}
\end{equation}
        \begin{equation}
\tau_{ik}[l] =\frac{1}{l!} \lim_{n \to \infty }
\frac{\partial^{\, l}}{\partial (n^{-2})^l}
\left( \frac{1}{n} \, \tau_{ik} (n \lambda, n m) \right).
\label{taul}
\end{equation}

     To obtain the explicit expression $\tau_{ik}[l]$
write the series for $S,U,V$ in reverse degrees of $n$
for $\lambda \to n \lambda, \ m \to n m $, \  $ n \to \infty $:
$S=\sum_{k=1}^\infty n^{-k} S_k, \ldots $
     Taking subsequent iterations in integral eqs.~(\ref{iuS}), (\ref{ie1})
and using the stationary phase method one obtains for first different
from zero terms
     \begin{equation}
V_1=W \,,\ \ U_2=DW \,,\ \ S_2=\frac{1}{4} W^2 \,, \ \
V_3=\frac{1}{2} W^3 - D^2 W - \frac{\omega}{2}
D\left(\frac{q}{\omega^3} \right),
\label{V1U2S2}
\end{equation}
           \begin{equation}
U_4=\frac{3}{2} \, W^2 DW - D^3 W - D \left( \frac{\omega}{2} \, D
\biggl( \frac{q}{\omega^3} \biggr) \right) + \frac{q}{2\omega^2} DW \,,
\label{U4}
\end{equation}
           \begin{equation}
S_4=\frac{3}{16} \, W^4 + \frac{1}{4}\,(D W)^2 - \frac{1}{2}\, W
D^2 W -\frac{1}{4}\,\omega W D\biggl(\frac{q}{\omega^3}\biggr) \,,
\label{S4}
\end{equation}
    where
\begin{equation}
q = \!\left( \Delta \xi R - \zeta R_{GB}^{\, 2} \right)\! a^2 , \ \
\omega=(m^2a^2+\lambda^2)^{1/2} , \ \ W=\frac{\omega\,'}{2 \omega^2} \,,
\ \ D=\frac{1}{2 \omega} \, \frac{d}{d\eta} \,.
\label{qoWD}
\end{equation}
     Note that in (\ref{V1U2S2})--(\ref{S4}) terms nonlocal on time
(dependent on $\eta$ and $\eta_0$) are excluded.
    These terms are absent if
$V_1(\eta_0) = V_3(\eta_0) = U_2(\eta_0) = U_4(\eta_0) = 0 $
which is supposed further.
    In particular, nonlocal terms are absent if first $2[N/2]$ derivatives
of the scale factor $a(\eta)$ of metric are zero in the initial moment of
time.

    Using~(\ref{ts00}), (\ref{tsab}), (\ref{V1U2S2})--(\ref{S4}) for
$\tau_{ik}[l]$ one obtains
     \begin{equation}
\tau_{00}[0] = \frac{\omega}{2} \,, \ \ \ \
\tau_{\alpha \beta}[0] = \delta_{\alpha \beta}\,
\frac{\lambda^2}{2 (N-1) \, \omega} \ ,
\label{at00}
\end{equation}
     \begin{equation}
\tau_{00}[1] = \omega S_2 + (N\!-\!1) \left( \Delta \xi -
\tilde{\zeta} c^2 \right) c V_1 +
\frac{N\!-\!1}{8 \omega} \biggl[ \,\Delta \xi \, 2 (N\!-\!2) c^2 +
\tilde{\zeta} (4\!-\!3N) c^4 \biggr] ,
\label{t001}
\end{equation}
                      \begin{eqnarray}
\tau_{\alpha \beta}[1] &=& \delta_{\alpha \beta} \left\{
\frac{1}{(N\!-\!1)\omega} \! \left[ \lambda^2 S_2 - \frac{m^2 a^2}{2}
\biggl( U_2 +\frac{q}{2 \omega^2} \biggr) \right]
\right. +
     \nonumber     \\
&+& \biggl[ \Delta \xi (N\!-\!1) - \tilde{\zeta} \biggl(
(N\!+\!1) c^2 - 2 c'\biggr) \biggr] c V_1 -
2 \left( \Delta \xi - \tilde{\zeta} c^2 \right) \omega U_2 +
\label{tab1}        \\
&+& \left. \frac{1}{4\omega} \Biggl[\, \Delta \xi (N\!-\!2)
\left( c^2 - 2c'\right) + \tilde{\zeta} c^2 (3N\!-\!4) \biggl(
2c' - \frac{3}{2} c^2 \biggr) \Biggr] \right\},
\nonumber
\end{eqnarray}
             \begin{eqnarray}
\tau_{00}[2] = \omega\, \biggl( S_4+\frac{q}{4 \omega^2}\,U_2 +
\frac{q^2}{16 \omega^4} \biggr) +
(N\!-\!1) \biggl( \Delta \xi - \tilde{\zeta} c^2 \biggr) c V_3 +
\nonumber    \\
{} +  \frac{N \!-\! 1}{4\, \omega} \biggl[ \, \Delta \xi \,
2 (N\!-\!2) c^2 - \tilde{\zeta} (3 N \!-\! 4) c^4 \biggr]
\biggl( S_2+\frac{1}{2} U_2+\frac{q}{4\omega^2} \biggr) \,,
\label{t002}           \\
\phantom{xxxxxxx}     \nonumber
\end{eqnarray}
                      \begin{eqnarray}
\tau_{\alpha \beta}[2] = \delta_{\alpha \beta} \left\{ \!
\frac{1}{N \!-\! 1} \Biggl[ \frac{\lambda^2}{\omega} \Biggl( S_4 +
\frac{q U_2}{4 \omega^2} + \frac{q^2}{16 \omega^4} \Biggr) \! -
\frac{m^2 a^2}{2 \,\omega}  \Biggl( U_4 +\frac{q^2}{4 \omega^4} +
\frac{q S_2}{\omega^2} \Biggr) \Biggr] + \right.
      \nonumber   \\
{}+ \biggl[ \Delta \xi (N \!-\!1) - \tilde{\zeta} \biggl(
(N\!+\!1) c^2 - 2 c'\biggr) \biggr] c V_3
- \left( \Delta \xi -  \tilde{\zeta} c^2 \right)
\Biggl( 2 \omega U_4 \!- \frac{q U_2}{\omega} \Biggr) +{}
\label{tab2}       \\
{} + \left. \Biggl[ \Delta \xi (N\!-\!2) ( c^2 \!-\! 2c')
+ \tilde{\zeta} (3N\!-\!4) c^2 \biggl( 2c' - \frac{3}{2} c^2 \biggr)
\Biggr] \frac{1}{2\omega}
\Biggl( S_2+\frac{1}{2} U_2+\frac{q}{4\omega^2} \Biggr) \right\}.
\nonumber
\end{eqnarray}
    These expressions give all information on subtractions for $N=4,5$.
    New counterterms appear for $N\ge 6$.
    For conformal scalar field they are given in Ref.~\cite{Pv2}.
    The renormalized  due to ~(\ref{Trik}) vacuum EMT is covariantly
conserved.
    This is seen from equations
$\nabla^i (\tau_{ik}/a^{N-2})=0 $ and
$\nabla^i (\tau_{ik}[l]/a^{N-2})=0 $ following from~(\ref{sdu}),
(\ref{ts00}), (\ref{tsab}), (\ref{at00})--(\ref{tab2}).

    To see the geometrical structure of the  $n$-wave procedure
counterterms let us make as in~\cite{MMSH} the  dimensional regularization.
    For calculation of integrals in dimensionally regularized counterterms
    \begin{equation}
T_{ik, \varepsilon }[l]=\frac{B_N}{a^{N-2}} (M)^{2 \varepsilon}
\int_{0}^{\infty} \! \lambda^{N-2} \tau_{ik, \varepsilon}[l]\, d\lambda\,,
\label{kTik}
\end{equation}
    with $\tau_{ik,\varepsilon}[l] $  defined by
(\ref{at00})--(\ref{tab2}) with putting $N \to N_0 - 2\varepsilon $,
one uses the equality
    \begin{equation}
\int_0^\infty  x^k\, (1+x^2)^{-p}\, dx =\frac{\Gamma\,(
\frac{k+1}{2})\,\Gamma\,(p-\frac{k+1}{2})}{2\, \Gamma\,(p)} \,.
\label{iGf}
\end{equation}
    In cases when the integral in the left hand side of~(\ref{iGf})
does not exist in usual sense it is taken as the analytic continuation of
the right side of~(\ref{iGf}) for corresponding values of $k$ and $p$.
    As the result of calculations one obtains
expressions~(\ref{TE0}), (\ref{TE1kp}), (\ref{TE2kp})
for the 0-th, 1-st and 2-nd counterterms $n$-wave procedure.
    So the geometrical structure of first three subtractions in the
$n$-wave procedure and in the effective action method occurs
to be the same.

    In conclusion note that Gauss-Bonnet coupling terms $R_{GB}^{\,2}$
can play important role in the early Universe.
    Effects of the nonzero value of $\zeta$ for scalar field can be
observed in black hole radiation,
in parameters of the so called bosonic stars~\cite{MielkeS}.
    Explicit value of $\zeta$ can be checked by experiment.

\vspace{11pt}
\section{Superheavy particles in the early Universe}

   It is known~\cite{GMM,Grib,GD} that the number of particles with mass
of the order of the Grand Unification scale created by gravitation in
the early Universe described by the radiation dominated Friedmann metric
is of the Dirac-Eddington order, i.e. of the observable order for the
visible mass.
   On the other side it is clear that if superheavy particles after
their creation continued to be stable for large enough time they will
lead to the  collapse of the Universe governed by the radiation
dominated  metric in the short on the cosmological scale time for
closed  Friedmann space or lead to the unrealistic scale factor for
the open space.
   So the idea was proposed that these superheavy particles
must decay on quarks and leptons with $CP$-noninvariance leading to
the observable baryon charge of the Universe before the time when
the energy density of the created superheavy particles will become
equal to that creating the background metric.
    If superheavy particles have nonzero baryon charge then their
decay in analogy with decay of neutral $K$-mesons will go as decay of
some short living and long living components.
   Supposing that  the lifetime of long living components is of
the cosmological order but their number was diminished in comparison
with the number of the short living components due to their interaction
with the baryon charge created previously similar to the well known
regeneration mechanism for $K$-mesons one can speculate about their
existence today as cold dark matter.
   Rare events of their decays can be identified as experimental
observations of high energetic cosmic rays~\cite{Takeda} with
the energy higher than the Greizen-Zatsepin-Kuzmin limit~\cite{GZK}.
    Here we shall discuss different possibilities of the role of
superheavy particles with the mass of the Grand Unification scale in
the early Universe.

   1) It is natural to think that some inflation era took place before
the Friedmann stage.
   Some inflaton field probably manifesting
itself as the quintessence in the modern epoch after the quasi de~Sitter
stage led to the dust like or to the radiation dominated
Friedmann Universe.
    Usually it is supposed that the inflaton field does not interact
with ordinary particles and can be some manifestation of
the non Einsteinian gravity for example due to high order corrections.
   So even if it decayed on some light ``inflaton'' particles
the primordial inflaton field can form hot dark matter but not the
visible matter and entropy present in background radiation.
    Our idea is that inflaton field was the source of Friedmann metric
with some small inhomogeneities, but visible matter and the
entropy of the Universe were created not by the inflaton field
itself but by the gravitation of this inflaton field.
   That gravitation created pairs of superheavy particles.
   Short living components decayed in time of the Grand Unification scale
and led to the nonzero baryon charge observed today as visible matter.
   If long living components had the lifetime of the order of
the ``early recombination era'' then the  energy density of created
long living particles soon became equal to that of the background inflaton
field (hot dark matter).
   Then the decay of all long living components led to the observable
entropy of the Universe.
   Here it is supposed that the energy density of the inflaton  field
led to the observed cosmological scale factor, so it is evident that
the created entropy due to our mechanism will be of the observable order.

   2) The other possibility is to put the hypothesis discussed by us
earlier~\cite{GrPv} that not all long living components decayed and
formed the entropy but some part of them survived up to modern time as
cold dark matter and superheavy particles  are  observed in
cosmic rays events.
   Then it is natural to suppose that the lifetime of the long living
component is of the cosmological order but the large part of them
regenerated into short living components  due to interaction with
the baryon charge in time shorter or equal to that of the
``early recombination era'' and entropy appeared due to this decay.

    Now let us give some numerical estimates.
     Total number of massive particles created in
Friedmann radiation dominated Universe
(scale factor $a(t)=a_0\, t^{1/2}$)      inside the horizon is as it
is known~\cite{GMM1,GMM}:
    \begin{equation}
N=n^{(s)}(t)\,a^3(t)=b^{(s)}\,M^{3/2}\,a_0^3 \ ,
\label{NbM}
\end{equation}
   where $b^{(0)} \approx 5.3 \cdot 10^{-4}$ for scalar
and  $b^{(1/2)} \approx 3.9 \cdot 10^{-3}$ for spinor particles
($ N \sim 10^{80} $ for $ M \sim 10^{14} $\,GeV, see Ref.~\cite{GMM}).
    Radiation dominance in the end of inflation era dark matter is
important for our calculations.
    If it is dust like the results will be different (see further).
    For the time ${t \gg M^{-1}} $ there is an era of going from the
radiation dominated model to the dust model of superheavy particles
    \begin{equation}
t_X\approx \left(\frac{3}{64 \pi \, b^{(s)}}\right)^2
\left(\frac{M_{Pl}}{M}\right)^4 \frac{1}{M}  \,.
\end{equation}
    If $M \sim 10^{14} $\,GeV,
$\ t_X \sim 10^{-15} $\,s for scalar and
$\ t_X \sim 10^{-17} $\,s for spinor particles.
   Let us call $t_X$ -- ``early recombination era''.
   For dust like ending of inflation era one has
$ N \sim M $ (see Ref.~\cite{KK}) and therefore the ratio of
the $X$-particles energy density $\varepsilon_X$ to the critical
density $\varepsilon_{crit}$  does not depend on time
($\varepsilon_X < \varepsilon_{crit}$ for $M < M_{Pl}$).

   Let us define $d $ --- the permitted part of long living
$X$-particles --- from the condition: on the moment of
recombination $t_{rec} $ in the observable Universe one has
$
d\,\varepsilon_X(t_{rec}) =\varepsilon_{crit}(t_{rec})  \,.
$
    It leads to
\begin{equation}
d=\frac{3}{64 \pi \, b^{(s)}}\left(\frac{M_{Pl}}{M}\right)^2
\frac{1}{\sqrt{M\,t_{rec}}}\, .
\label{d}
\end{equation}
   For $M=10^{13} - 10^{14} $\,GeV one has
$d \approx 10^{-12} - 10^{-14} $ for scalar and
$d \approx 10^{-13} - 10^{-15} $ for spinor particles.
     So the life time of main part or all $X$-particles must be smaller
or equal than $t_X$.

    Now let us construct the model which can give: \
a) short living $X$-particles decay in time
   $\tau_q < t_X $ (more wishful is
   $\tau_q \sim t_C \approx 10^{-38} - 10^{-35} $\,s,
i.e. Compton time for $X$-particles) \
b) long living particles decay with $\tau_l \approx t_X $.
   Baryon charge nonconservation with $CP$-nonconservation in full
analogy with the $K^0$-meson theory with nonconserved hypercharge and
$CP$-nonconservation leads to the effective Hamiltonian of the decaying
$X, \bar{X}$ - particles with nonhermitean matrix.

  For the matrix of the effective Hamiltonian
$ H=\{ H_{ij} \}, \ {i,j=1,2}$  let $H_{11}\! =\! H_{22}$
due to $CPT$-invariance.
    Denote
$\ \varepsilon=(\sqrt{\vphantom{ }H_{12}} - \sqrt{H_{21}}\,)\, / \,
(\sqrt{H_{12}} + \sqrt{H_{21}} \, )$.
    The eigenvalues $\lambda_{1,2} $ and eigenvectors
$|\Psi_{1,2}\rangle $  of matrix $H$ are
    \begin{equation}
\lambda_{1,2} = H_{11} \pm \frac{H_{12}+H_{21}}{2} \,
\frac{1-\varepsilon^2}{1+\varepsilon^2} \,,
\end{equation}
    \begin{equation}
|\Psi_{1,2}\rangle =\frac{1}{\sqrt{2\,(1+|\varepsilon |^2)}}\,
\left[ (1+\varepsilon) \,|1\rangle \pm \,(1- \varepsilon) \,
 |2\rangle \right].
\end{equation}
         In particular
\begin{equation}
H=     \left(
\begin{array}{cc}
E-\frac{i}{4}\left(\tau_q^{-1} +\tau_l^{-1}\right)
  &
\frac{1+\varepsilon}{1-\varepsilon}
\left[A-\frac{i}{4}\left(\tau_q^{-1} -\tau_l^{-1}\right)\right]
 \\  & \\
\frac{1-\varepsilon}{1+\varepsilon}
\left[A-\frac{i}{4}\left(\tau_q^{-1} -\tau_l^{-1}\right)\right]
 &
E-\frac{i}{4}\left(\tau_q^{-1} +\tau_l^{-1}\right) \\
\end{array}        \right) .
\label{HM}
\end{equation}

    Then the state $|\Psi_1 \rangle $ describes
short living particles $X_q$ with the life time
$ \ \tau_q \ $ and mass $E+A$.
    The state $\ |\Psi_2 \rangle $ is the state of long living particles
$X_l$ with life time $ \tau_l \ $  and mass $E-A$.
    Here $A$ is the arbitrary parameter $-E<A<E$  and it can be zero,
$E=M$.

    So for the scenario~1) it is sufficient to take
$\tau_l \approx t_X$.

    In scenario~2)
the small $ d \sim 10^{-15} - 10^{-12} $ part of long living
   $X$-particles with $\tau_l > t_U \approx 4.3 \cdot 10^{17}$\,s
   \  ($t_U $ is the age of the Universe)
is forming the dark matter.
   The decay of these superheavy particles in modern epoch
can give observed ultra high energy cosmic rays.
    Using the estimate for the velocity of change of the concentration of
long living superheavy particles~\cite{BBV} \
$|\dot{n}_x| \sim 10^{-42}\, \mbox{cm}^{-3}\,\mbox{s}^{-1} $,
and taking the life time $\tau_l $ of long living particles as
$2\cdot 10^{22} $\,s, we obtain concentration
$n_X \approx 2\cdot 10^{-20} \,\mbox{cm}^{-3} $ at the modern epoch,
corresponding to the critical density for $M=10^{14} $\,GeV\,.

   Let us use the model with effective Hamiltonian~(\ref{HM})
where $\tau_l > t_U$ and take into account
transformations of the long living component into the short living
one due to the presence of baryon substance created by decays of
the short living particles in analogy with the regeneration
mechanism for $K^0$-mesons.

    Let us investigate the model with the interaction which in the
basis   $\ |1 \rangle, \ |2 \rangle $  is described by the matrix
     \begin{equation}
H^d =     \left(
\begin{array}{cc}
0  & 0   \\
0  & - i \gamma \\
\end{array}        \right).
\label{Hd}
\end{equation}
    The eigenvalues of the Hamiltonian  $H+H^d$  are
     \begin{equation}
\lambda^d_{1,2} = E - \frac{i}{4}
\left(\tau_q^{-1} + \tau_l^{-1} \right) -i\,\frac{\gamma}{2} \pm
\sqrt{ \left( A - \frac{i}{4} \left(\tau_q^{-1} - \tau_l^{-1} \right)
\right)^2 -\frac{\gamma^2}{4} } \ .
\label{lamdop}
\end{equation}
    In case  when   $\gamma \ll \tau_q^{-1}$
for the long living component one obtains
     \begin{equation}
\lambda^d_{2} \approx  E - A - \frac{i}{2}\, \tau_l^{-1}
-i\,\frac{\gamma}{2} \,,
\label{ldolg}
\end{equation}
     \begin{equation}
\| \Psi_2(t) \|{}^2 = \| \Psi_2(t_0) \|{}^2 \exp \left[
\frac{t_0 - t}{\tau_l} - \int_{t_0}^t \gamma(t)\, d t \right].
\label{P21}
\end{equation}

    The parameter $\gamma$,  describing the interaction with the
substance of the baryon medium, is evidently dependent on its state
and concentration of particles in it.
   For approximate evaluations take this parameter as
proportional to the concentration of particles:
$\gamma = \alpha\, n^{(0)}(t)$.
       Putting
$\tau_l = 2 \cdot 10^{22}$\,s, \ $t \le t_U$, \ $a(t)=a_0 \sqrt{t}$\
by~(\ref{NbM})   one obtains
     \begin{equation}
\| \Psi_2(t) \|^2 = \| \Psi_2(t_0) \|^2 \exp \left[ \alpha 2 b^{(s)}
M^{3/2}\left( \frac{1}{\sqrt{t}} - \frac{1}{\sqrt{t_0}}
\right) \right].
\label{Ptt0}
\end{equation}
      So the decay of the long living component due to this mechanism
takes place close to the time  $t_0$.
    One can think that this interaction of $X_l$ with baryon charge
is effective for times, when the baryon charge becomes strictly
conserved, i.e. we take the time larger or equal to the electroweak
time scale, defined by the temperature of the products of decay
of $X_q$.
    This temperature is defined from
$ M n^{(s)}(\tau_q) \approx \sigma T^4 $
and is given by

     \begin{equation}
T(t) = \left( \frac{ 30\, b^{(s)} }{ \pi^2 N_l } \right)^{\!1/4}
\frac{ M^{5/8}\, \tau_q^{1/8} }{ k_B\, \sqrt{t} },
\label{T}
\end{equation}
   where $k_B$ is Boltzmann constant, $N_l$ is defined by the number of
boson $N_B$ and fermion $N_F$ degrees of freedom of all kinds of
light particles:
$N_l=N_B + \frac{7}{8} N_F$\, (see Ref.~\cite{KKZ}).
   At time $t_X$ this temperature is equal to
     \begin{equation}
T(t_X) = \frac{64 \sqrt{\pi}}{3}
\left( \frac{ 30}{N_l} \right)^{\!1/4}\!\!
\left( b^{(s)} \right)^{\!5/4}\!
\left( M \tau_q \right)^{1/8}  \frac{ M^3}{k_B M_{Pl}^2}\,.
\label{TtX}
\end{equation}
   If $ \tau_q = 1/M $ and $N_l\sim 10^2$ -- $10^4 $, then for spinor
$X$-particles $ T(t_X) \approx 300$ -- $100 $\,GeV,
i.e. the electroweak scale for created particles
(which is however different from that for the background).

    So let us suppose $t_0 \approx t_X$.
    If  $d$ -- is the part of long living particles surviving up to
the time  $t$   $\ (t_U \ge t \gg t_C)$  then from~(\ref{d})
and (\ref{Ptt0})
one obtains the evaluation for the parameter~$\alpha$
     \begin{equation}
\alpha = \frac{ - 3 \ln d}{ 128 \pi (b^{(s)})^2} \,
\frac{M_{Pl}^2}{M^4} \,.
\label{ald}
\end{equation}
    For  $M=10^{14}$\,GeV    and  $d=10^{-15}$   one obtains
$\alpha \approx 10^{-30}$\,cm${}^3$/s.
    If  $\tau_q \sim 10^{-38} - 10^{-35} $\,s
then the condition  $\gamma(t) \ll \tau_q^{-1} $
used in Eq.~(\ref{ldolg}) is valid for $t> t_X$.
    For this value  $\alpha$ we have
$\gamma(t_U)\approx 10^{-36}$\,s${}^{-1}$ $\ll \tau_l^{-1}$.
   So one can neglect this mechanism for the decay of the long living
component of $X$-particles for the modern epoch
while for early universe at $t_0 \approx t_X$ it was important.
    The observable entropy in this scenario is created due to decay of
$X_l$ on quarks and antiquarks at the time $t_X$ when the Grand
Unification symmetry is totally broken.
    Baryon charge is created at $t_q $ which can be equal to
Compton time for $X$-particles $t_C \sim 10^{-38} - 10^{-35} $\,s.

   Our scheme is the same for the scalar particles and the fermions.
   The superheavy fermions are used, for example, in some models of
neutrino mass generation (the {\it see-saw} mechanism) in
Grand Unification theories~\cite{GMRS,Yosh}.
    New experiments on high energetic particles in cosmic rays
surely will give us more information on their structure and origin.

\vspace{1mm}
{\it Acknowledgements}.
This work was supported by Ministry of Education of Russia, grant
E02-3.1-198.

\vspace{14pt}

\end{document}